# First-principles investigation of electron-induced cross-linking of aromatic self-assembled monolayers on Au(111)

Pepa Cabrera-Sanfelix,[1] Andrés Arnau[1,2,3] and Daniel Sánchez-Portal [1,2]

[1] Donostia International Physics Center (DIPC), Paseo Manuel de Lardizabal 4, San Sebastian 20018, Spain
[2] Centro de Física de Materiales CFM-MPC, Centro Mixto CSIC-UPV/EHU, Apdo. 1072, San Sebastián 20080, Spain
[3] Departamento de Física de Materiales UPV/EHU, Facultad de Química, Apdo. 1072, San Sebastián 20080, Spain

We have performed a density functional theory study of the possible layered geometries occurring after dehydrogenation of a self-assembled monolayer (SAM) of biphenyl-thiol molecules (BPTs) adsorbed on a Au(111), as it has been experimentally observed for low energy electron irradiated SAMs of 4'-nitro-1,1'-biphenyl-thiol adsorbed on a Au(111) surface. [*Eck, W. et al., Advanced Materials* **2000**, *12, 805*] Cross-link formation between the BPT molecules has been analyzed using different models with different degrees of complexity. We start by analyzing the bonding between biphenyl (BP) molecules in a lineal dimer and their characteristic vibration frequencies. Next, we consider the most stable cross-linked structures formed in an extended free-standing monolayer of fully dehydrogenated BP molecules. Finally, we analyze a more realistic model where the role of the Au(111) substrate and sulphur head groups is explicitly taken into account. In this more complex model, the dehydrogenated BPT molecules are found to interact covalently to spontaneously form "graphene-like" nanoflakes. We propose that these nanographenes provide plausible building-blocks for the structure of the carbon layers formed by electron irradiation of BPT-SAMs. In particular, it is quite tempting to visualize those structures as the result of the cross-link and entanglement of such graphene nanoflakes.

## 1. Introduction

Self-assembled monolayers (SAMs) of organic species adsorbed on metal and semiconductor surfaces have attracted great interest during the past decades due to their technological applications, such as surface lubricant and wetting, micro- and nano-lithography, biomedical, optical and microelectronic devices.[1-5] SAMs are formed by three main parts: a *head group* that binds to the substrate; a *tail group* that constitutes the outer surface of the film; and the *spacer* that connects the head with the tail groups. The spacer essentially controls the intermolecular separation and the degree of order in the film. A large part of the interest has focussed on aromatic moieties as an alternative of alkane-based spacers.[6-11] Aromatic SAMs differ significantly from the aliphatic ones in geometry and intermolecular interactions, so they show different charge-transfer properties[12-14] and different behaviour during electron irradiation.[15,16]

Here, we are interested in SAMs containing spacers formed by biphenyl (BP) groups, and, in particular, in the modifications of the structures of these films after the dehydrogenization process induced by electron irradiation. Recent experiments along these lines have been performed for 4'-nitro-1,1'-biphenyl-thiol (BPT) adsorbed on Au(111) surface and other related SAMs. Scaning Tunelling Microscopy (STM) images of this film show a well defined, densely packed, $2\sqrt{3}\times\sqrt{3}$ superstructure where the molecules are oriented almost upright, with respect to the gold substrate.[17] A similar adsorption geometry is found for other aromatic SAMs, like the so-called BPn with n an odd integer, which consist of two phenyl rings as a spacer group in addition to an alkene-thiolate chain $(CH_2)_n$-SH as a head group.[18-20] Several experiments have shown that continuous films of cross-linked molecules are formed after low energy (~50 eV) electron irradiation of these type of SAMs. Evidence of lateral cross-linking of the BPT units, accompanying the electron induced cleavage of the C-H bonds, has been obtained with a variety of techniques like photoelectron (XPS) and infrared (IR) spectroscopies, as well as X-ray adsorption (NEXAFS).[5, 8, 9, 11, 15, 16, 21, 22]

The cross-links increase the stability and mechanical properties of the irradiated regions, so that irradiated BPT-SAMs could be used as an ultrathin (~15Å) negative resist for nanolithography.[15] Furthermore, these SAMs are mechanically stable as suspended monolayers even at temperatures above 1200K. This high thermal stability makes them suitable candidates for applications as lubricants in mechanical contacts.[10] Other properties of the freely standing nanosheets, as the conductivity and stiffness, can be controlled and could also be exploited in applications.[5]

In this work, we present a first attempt to theoretically describe the geometry and stability of the cross-linked structures present in the irradiated SAMs containing BP units as spacers. Our simulations indicate that very different structures are obtained depending on (i) whether the tail and head groups are preserved during irradiation (or at least during the initial cross-link formation), and (ii) the density of the film, which in turn depends on the substrate. Of course, the approximations that we need to make are quite severe and many questions still need further work. A few relevant ones are: what is the yield and the dynamics of the C-H bond cleavage process as a function of the electron energy?, how does it depend on the substrate, the tail and head groups?, what is the effect of temperature? and, in particular, what is the optimum energy dissipation rate, after the formation of each new C-C bond, for the formation of cross-linked structures?. Our simulations are just a first step towards the understanding of this complex process.

The results are presented in four different sections. First, we study the bonding and vibration frequencies in a very simple model: the BP dimer. Next, we present the most stable structures formed by the cross-link of dehydrogenated BPT units in a free-standing layer, i.e., without considering the presence of the substrate. Then, we explore the adsorption geometry of a BPT-SAM on Au(111) within our DFT methodology and, finally, we study the transformation that this structure undergoes after the dehydrogenation of the BPT molecules.

## 2. Method

Our study is based on first-principles calculations within the framework of Density Functional Theory (DFT). All calculations were performed using the Vienna ab initio simulation package (VASP),[23-25] employing a plane-wave basis set and implementing DFT within the Perdew-Wang 1991 (PW91) version of the general gradient approximation (GGA).[26] The projector augmented wave (PAW) method[27, 28] was used to describe the interaction of valence electrons with Au, S, C and H cores. For computational simplicity the theoretical BPT units consist of two phenyl rings as spacer and one S atom that strongly bonds to the substrate, as a head group. No tail group was considered since our interest focuses on the possible cross-liking between neighboring phenyl rings. In the particular case in which the substrate is not included, not only the atomic positions, but the size and shape of the unit cell are also relaxed, so that we can estimate the degree of compression of the free-standing BP-monolayers. The gold substrate has been represented by four layers of Au(111) and we have used a 2√3x2√3 supercell, with periodic boundary conditions to represent the experimentally found 2√3x√3 superstructure for the adsorbed BPT-SAM on Au (111).[17] Therefore, four BPT units can be accommodated in the simulation cell. A vacuum gap of 10.5Å along the normal to the slab was found to be large enough for avoiding interaction between periodic images.

In all calculations, we have used a 3x3x1 k-point sampling and a plane-wave cutoff of 20 Ry. All BPT units and the topmost layer of the Au(111) substrate were allowed to relax until the residual forces were less than 0.03 eV/Å. The convergence of our computations has been checked for the BP dimer. For this system increasing the plane-wave cutoff energy to 25 Ry and using a more stringent force convergence criteria of 0.01 eV/Å only changed the calculated total energy within 10 meV. The vibration frequencies for the different structures have been computed using the finite displacement method and only considering those atoms in the BP units (this approximation is well justify due to the large mass mismatch with the Au atoms of the substrate). For periodic structures only Gamma point phonons have been calculated and a Gaussian broadening of 1.5 cm$^{-1}$ has been used to plot the vibration density of states (VDOS). Due to the inherent limitations of the calculation methods and the many unknowns about the structure of the studied systems, we cannot expect a very good agreement with the measured experimental spectra. Thus, we focus in the shifts of the dominant peaks in the vibration spectra upon the formation of the cross-linked structures. Hopefully this information is linked to the formation of particular structural patterns and can be of interest beyond the particular models considered here.

## 3. Results

We start by exploring the consequences of cross-link formation in a BP dimer. In particular, we try to identify the vibration modes that could be used as a signature of cross-link formation. The problem of the formation and structure of free-standing cross-linked SAMs as a consequence of electron irradiation is a formidable one. It requires the simultaneous description of the cleavage by irradiation of C-H and Au-S bonds, as well, as the formation of new C-C bonds. This is well beyond the capabilities of current first-principles simulations due to both the complexity of the processes involved, and the large size of the simulation cell required. Therefore, we need to perform somewhat severe approximations in our study. Accordingly, we have decided to explore two simplified limiting cases:

(i) In a first series of simulations, we disregard the effect of the gold substrate and only consider the structures obtained when a free-standing BPT-SAM is completely dehydrogenated and both head (S) and tail (H) groups are also detached as a consequence of irradiation. Under these conditions, we obtain a largely compressed carbon nanosheet with a characteristic structure exhibiting eight-carbon rings.

(ii) In a second group of simulations we explicitly take into account the effects due to the presence of the substrate and consider a commensurate structure of a BPT-SAM adsorbed on Au(111). All C-H bonds are assumed to suffer cleavage due to irradiation. However, BPT units still keep their tail and thiolated head groups. Under these conditions, the biphenyl molecules in the unit cell tend to align and form graphene-like nanoflakes.

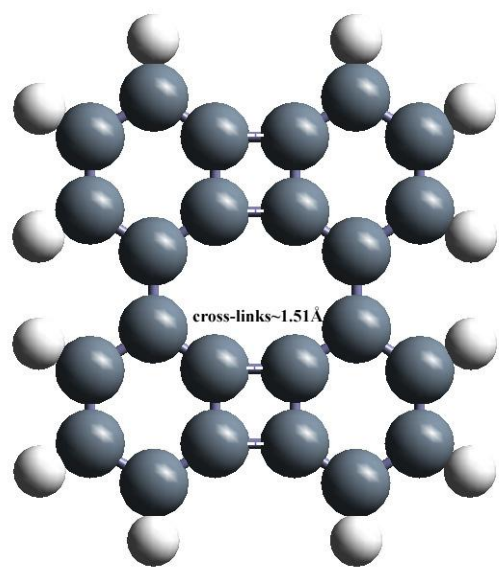

**Figure 1.** Dimer formed by two aligned biphenyl molecules connected by four crosslinks (marked in dashed white): two internal, forming part of the middle octagon, of 1.515 Å and two external of 1.506Å. Energy of ~2.778 eV is released by the formation of each cross-link.

**Cross-linked BP dimer**

As a first approximation to the cross-link formation on the dehydrogenated BPT-SAM, we have considered a dimer formed by two aligned and cross-linked biphenyls. The relaxed structured is illustrated in Figure 1. There are four links between the two BP molecules, all of them with bond lengths close to 1.5 Å. The intermolecular bonds give rise to the appearance of characteristic four-carbon rings with a square shape. The VDOS of a BP monomer and the dimer are presented in Figure 2 (a) and (b), respectively. Comparing these two vibration spectra we can identify some peaks related to the presence of cross-links in the dimer configuration. In order to facilitate this identification, we have highlighted the contribution from the carbon squares only present in the case of the dimer structure. The most significant differences appear in the high energy part of the spectra. For the BP monomer (Fig. 2 (a)) we have a well defined peak around 1590 cm$^{-1}$ associated with the stretching of the aromatic C-C bonds. For the BP dimer, Fig. 2 (b), we can see that this peak splits in two. One of the peaks shifts down slightly, appearing around 1585 cm$^{-1}$, and again corresponds mainly to stretching vibrations in the aromatic rings. However, a new peak appears at higher frequencies, around 1683 cm$^{-1}$, which is mainly associated with the stretching of the intermolecular bonds in the BP dimer.

The observed splitting and upwards shift of the stretching modes upon formation of cross-links between BP molecules is widely consistent with the experimental observation.[11] For example, for pristine BPT-SAMs on Au(111) a peak is observed at 1598 cm$^{-1}$ clearly associated with the aromaticity of the BPT units. After irradiation with electrons the peak moves to higher energies, 1611.5 cm$^{-1}$, and becomes much broader, featuring a clear shoulder at frequencies closer to those of the original signal.

Therefore, from our calculations we conclude that the breaking of aromaticity due to the formation of cross-links between BP molecules, with the appearance of characteristics four-carbon rings, is accompanied with a shift to higher energies of the corresponding C-C stretching modes. If similar structures are indeed present in the experimental samples, this shift could be use as an indication of the cross-link formation. We now proceed to consider more realistic models of the BPT-SAMs in which we try to identify similar fingerprints for the formation of cross-links. However, as we will see, the situation becomes more complicated since, in some cases, the structures depart considerably from that of the ideal BP dimer.

**Cross-linked BP-SAM in the absence of substrate**

In the experiment, the irradiation of BPT-SAMs on Au(111) causes the cross-linking of the BPTs units and, consequently, a significant compression of the structure. The resulting densely packed monolayer is highly stable against thermal treatments[10] and can be detached from the surface using different methods, thus providing an efficient route to obtain stable free-standing carbon nanosheets with promising technological applications.[5, 21] As a first attempt to find plausible candidates for the structure of these nanosheets, we have investigated simple models of cross-linked structures for completely dehydrogenated BPs in the absence of any substrate. As a starting guess for our geometries we use a molecular arrangement and density similar to those of the BPT-SAMs on Au (111).

Thus, we place two fully dehydrogenated BP molecules in a 2√3x√3 supercell of the Au(111) surface. However, in addition to the atomic positions, in these calculations we also relax the size and shape of the unit cell so that the density of the layer is optimized and we can estimate the degree of compression of the free-standing BP-monolayers.[5] The most stable configuration is presented in Figure 3, with the structural unit illustrated in the inset for a better visualization. The structure suffers a strong reorganization: not only there are new bonds formed between BP molecules, but the hexagons of one of the BP molecules open and form two characteristic bent octagonal-rings in the structure (marked in red in the inset of Figure 3). The areal density of the structure increases by ~3.8 % with this reorganization. As can be expected the driving force behind the observed reorganization is the creation of new C-C bonds. In particular, all the carbon atoms in the octagonal rings increase their coordination to become three-, or even, four-fold-coordinated as in the case of the atoms belonging to the square-rings.

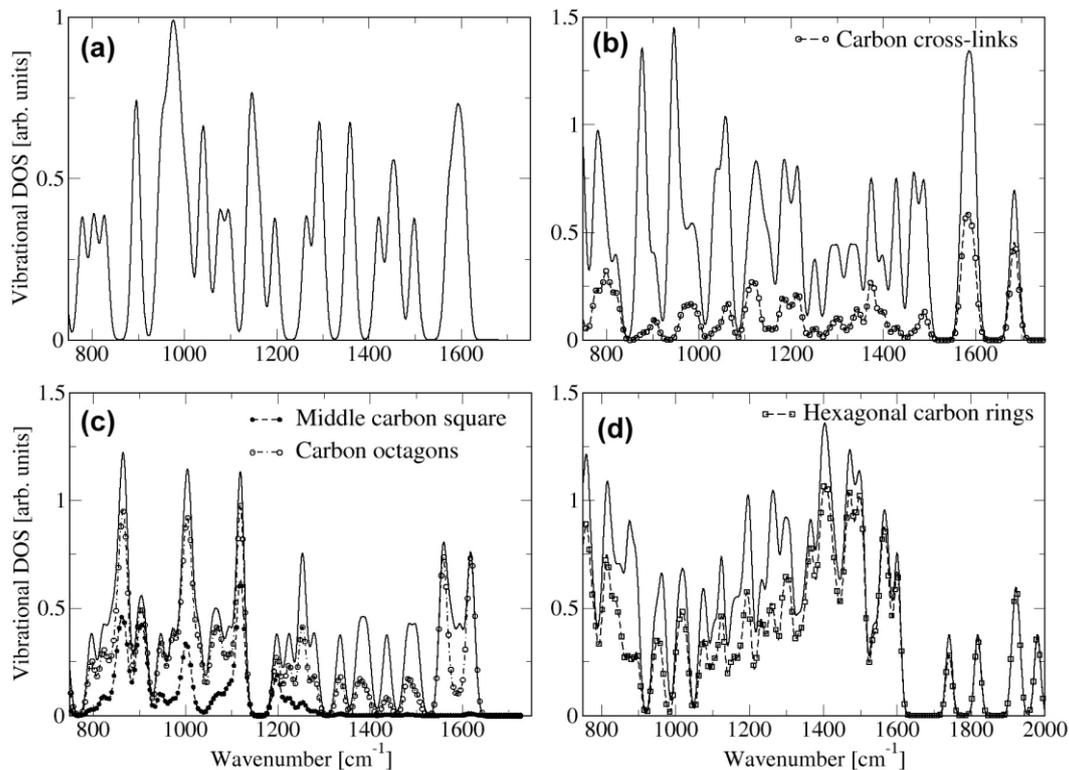

**Figure 2.** Calculated Vibration Density of States (VDOS) for (a) a BP-monomer, (b) a lineal BP-dimer, (structure shown in Figure.1), (c) a compressed free-standing cross-linked BP monolayer (see Figure 3), and (d) graphene nanoflakes on Au(111) (see Figure 5).

The VDOS at the Gamma point of this compressed free-standing BP monolayer is presented in Figure 2(c). Comparing with the case of the BP-monomer [Fig. 2 (a)], we observe a splitting of the highest energy peak: two well defined peaks appear now centred at 1618 cm$^{-1}$ and 1559 cm$^{-1}$. In Figure 2 (c) we have highlighted the contribution from the carbon atoms belonging to the octagonal rings. Taking into account this information, it seems that the observed splitting is associated with the formation of the octagonal carbon rings. This kind of splitting of the C-C stretching modes is again roughly consistent with the experimental IR information. Indeed the peak at 1618 cm$^{-1}$ is in very good correspondence with the experimentally measured signal close to 1611 cm$^{-1}$.

However, the position of the low energy peak does not agree with any clearly measured signal. Another important difference with the case of the BP dimer is that the carbon square (marked in blue on the inset of Figure 3) does not present any characteristic peak at high frequencies for the compressed BP monolayer. However, it contributes significantly to a strong peak appearing around 1119 cm$^{-1}$. The position of this peak is consistent with the four-fold coordination of these atoms mentioned above and, thus, the presence of single C-C bonds,[29, 30] but unfortunately it cannot be directly associated to any clear peak in the experimental IR spectra.

Therefore, we can conclude that the computed frequencies for our model of a free-standing cross-linked BP monolayer do not accurately reproduce the available experimental data. This seems to indicate that the real structures formed by irradiation of BPT-SAM on Au(111) depart significantly from that shown in Figure 3. This might not be surprising if we take into account the strong assumptions behind these simulations. In this first type of calculations we have considered monolayers solely formed by carbon, whereas experimentally there is clear evidence that only C – H cleavage occurs during irradiation, whereas the tail and head groups do not disappear but transform in irradiation induced species,[8, 11, 16] being removed in later processes.[5, 21, 22] Therefore, in our next model, we consider the Au(111) substrate explicitly and dehydrogenate the BPT units, but allow them to keep their head and tail groups.

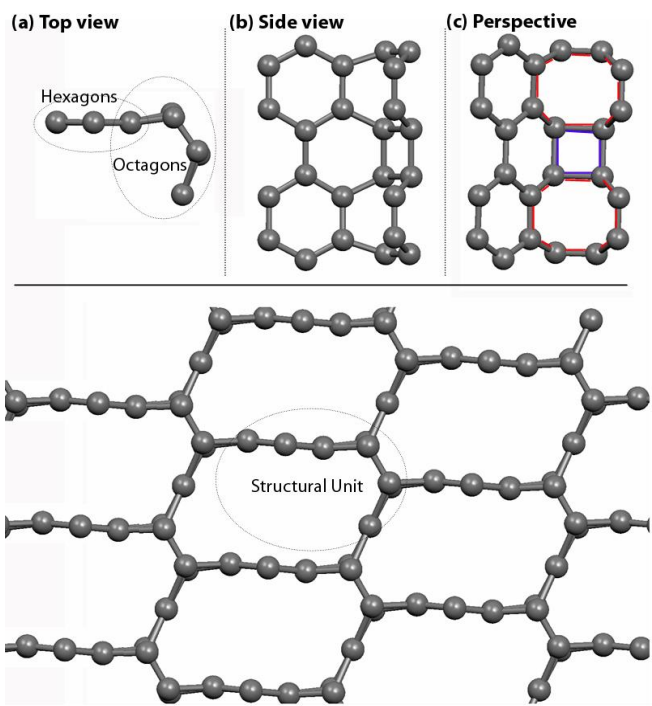

**Figure 3.** Lower panel: Top view of the most stable free-standing carbon nanosheet obtained after full relaxation (geometry and unit cell) of a system containing two dehydrogenated BP molecules per unit cell. We have estimated 12 cross-links per unit cell, with an estimated energy per cross-link of 2.14 eV. The cross-linked structural unit consists of two restructured BPs into three hexagons, two octagons (marked in red) and one square network (marked in blue). Upper panels: (a) the top view, (b) a side view and (c) a perspective of the structural unit.

**BPT-SAM on Au(111)**

An initial relaxation of the substrate using a 2√3x2√3-cell, was followed by the adsorption of four BPT units following a 2√3x√3 arrangement similar to that deduced from STM images.[19] In order to have the necessary commensuration with the underlying substrate, we need to apply a substantial (~6%) compression to the monolayer of BPTs respect to the experimental structure. Therefore, BPTs cannot be properly adsorbed on their preferred fcc sites. In fact, we find a slightly disrupted adsorption geometry. In the relaxed system three BPTs are adsorbed with the S atoms on fcc-like sites (slightly displaced from the symmetric fcc site), forming bonds with the three neighbouring Au atoms on the topmost layer and with different adsorption heights, 1.45Å, 1.79Å and 2.07Å (see Figure 4). The fourth BPT molecule in our structure gets adsorbed also at approximately fcc position but farther away from the surface, with its S head group at 2.81Å from the topmost layer (molecule labelled 1 in Fig.4). As a consequence, it pulls out one Au atom ~0.6Å from the Au(111) topmost layer. The typical differences between the adsorption heights of different BPT and BPn molecules in SAMs on Au(111) have been reported experimentally to be in the range from 0.2Å to 0.5Å[17, 18] from two well defined heights to four different levels. These observations are in reasonable agreement with three of the BPTs in our cell. However, the fourth one is adsorbed 1.3 Å higher than the lowest one. We think that this is principally due to the computational limitations of using a too small cell. Unfortunately, a supercell that provides a better commensuration between the BPT-SAM and the Au(111) substrate is prohibitively large to perform first-principles calculations. There is, however, another possibility to explain the disagreement with the experimentally observed structure. It has been recently demonstrated that many organosulfur molecules prefer to bond to Au adatoms present in the surface.[31, 32] This happens because Au adatoms favour the dissociation of the S-H bond and formation of $H_2$ during adsorption.[32] It might be that, even using our simulation cell, the adsorption of the BPT units on top of Au adatoms would result in a relaxed structure with a less disrupted gold surface. However, there have also been claims that when molecules adsorb on adatoms the resulting films show a less order structure. In contrast, the initial structure of the SAMs that we consider here is known to have a well-defined periodicity. Thus, the possibility of the adsorption on Au adatoms will not be pursued here and we leave this line of research for future work.

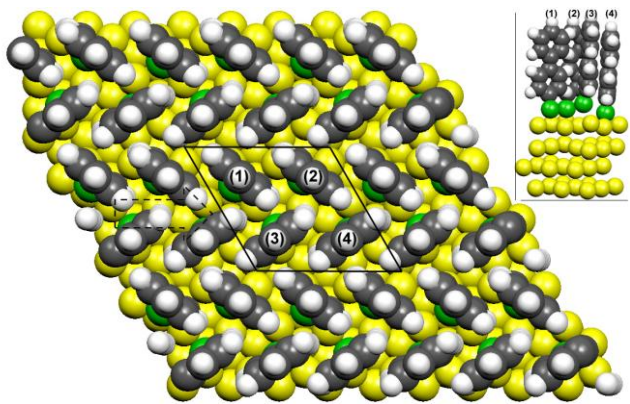

**Figure 4**. Large Image: 3x3 top-visualization of the DFT-optimized BPT-SAM adsorbed on Au(111) surface following the 2√3x√3 superstructure. Notice that the cell used for this calculation is 2√3x2√3, and we have placed 4 BPT units in it. Inset: side view of the surface unit cell area (rhomboid marked by the dashed arrow). Adsorption takes place close to fcc sites at different heights: molecule (1) at 2.07 Å, (2) at 1.79 Å, (3) at 2.81 Å, and molecule (4) at 1.45 Å.

Anyhow, despite the restrictions imposed by the simulation cell, the monolayer is well adsorbed on the substrate with an average adsorption energy of 1.13 eV per BPT unit. Therefore, we believe that this structure may be taken as a reasonable initial geometry to study the cross-link of dehydrogenated BPT molecules adsorbed on Au(111).

**Cross-linked SAM on Au(111)**

Cross-linking of BPTs occurring from the 2√3x√3 superstructure adsorbed the Au(111) substrate represents a more realistic scenario to compare with the experiment. Furthermore, there is experimental evidence that C-H cleavage takes place before any variation or desorption of the tail groups.[8, 11, 16] Therefore, in this case we only consider the cleavage of C-H bonds and leave the head and tail groups of the BPTs. Here, we use a very simplified model in which the head group of the BP units is formed by just one H atom. Figure 5 shows the optimized configuration. Some of the molecules have rotated and significantly changed their positions, and the four BPT units in the unit cell have coalesced to form a small "graphene-like" nanoflake. The top and bottom edges of these graphene nanoflakes are saturated with hydrogen and sulphur, respectively. The differences between the heights of the sulphur atoms over the substrate are now more pronounced. Two S atoms adsorb close to fcc sites at 1.54 Å and 1.90 Å over the Au topmost layer, whereas the other two S atoms rise to 3.08 Å and 3.85 Å from the surface (see Figure 5). Notice that the Au atom that was previously pulled out from the surface has returned now to the topmost layer level.

Despite the large S-Au average distance, the adsorption energy of each graphene nanoflake with the gold substrate remains as large as ~2.3 eV. This explains the experimental thermal stability of the cross-linked aromatic self-assembled monolayers on Au surfaces.[10]

The lateral edges of these nano-graphenes remain unsaturated and, therefore, tend to form covalent bonds with neighbouring flakes. In the present calculations this is not possible due to the periodic boundary conditions that fix the density of the layer and keeps the flakes separated from each other by at least 4 Å on the xy-plane. However, in the experiment we can expect these nanoflakes to entangle and form an extended two-dimensional structure with higher areal density. This seems to be in agreement with the experimental observation of dense self-assembled monolayers after irradiation of BPT-SAMs adsorbed on Au(111).[10, 15, 22]

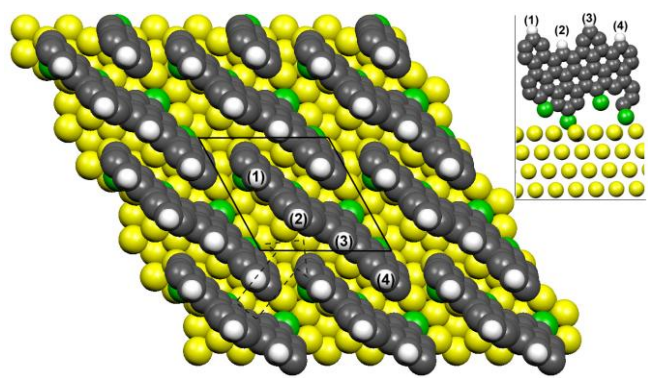

**Figure 5**. Large Image: Optimized geometry of the structure shown in Figure 4 after dehydrogenization of the four BPT molecules. Inset: side view of one carbon nanoflake. Sulphur atoms of molecules (1) and (3) are 3.85 Å and 3.08 Å over the surface, respectively. Molecules (2) and (4) adsorb at fcc sites with heights of 1.54Å and 1.90Å.

We have computed the vibration frequencies of the supported graphene flakes shown in Figure 5. We have not considered the movements of the substrate atoms in the calculations, but uniquely those of the atoms forming the nanoflakes. The VDOS is shown in Figure 2(d). Again we observe a splitting of the highest energy peak appearing for the BP-monomer (Figure 2(a)) into two peaks at 1603cm$^{-1}$ and 1565cm$^{-1}$. The red curve corresponds to the contribution of the carbon atoms forming aromatic rings, but not bonded to any tail or head group. So the split peak is due to the aromatic stretching and, therefore, corresponds with the experimental IR peak found at 1611.5 cm$^{-1}$ for the irradiated BPT-monolayer on Au(111).[11] Other well differentiated peaks are found at higher energies, 1740, 1818, 1923 and 1981 cm$^{-1}$, although they cannot be directly correlated with the available experimental information. These high energy peaks are related to the low-coordinated carbon atoms appearing in the lateral edges of the structure.[29, 30]

Since our periodic boundary conditions avoid the cross-link between flakes and, therefore, the formation of extended structures, it is not surprising that the agreement between our calculated vibrations and the experiment is limited. Despite our computational limitations, our results suggest that a structure composed of cross-linked graphene nanoflakes is a very plausible geometry for irradiated and, therefore, dehydrogenated and cross-linked BPT-self-assembled, monolayers on Au(111).

**Summary and Conclusions**

We have performed DFT calculations to investigate the possible structure of the cross-linked BPT layered structures experimentally obtained after electron irradiation of BPT SAMs on metal surfaces.[5, 10, 11, 21, 22] As a first approximation, we optimized the geometry of a lineal BP dimer and calculated the vibration modes of the formed cross-linked structure. This gives us a first hint to try an identification of the absorbance peaks observed in the IR spectra of cross-linked BPT self-assembled monolayers.[11] Then, a simple model of the cross-linked structure for completely dehydrogenated BPs was proposed in absence of the gold substrate. This structure is very favourable energetically and is characterized by the appearance of carbon octagonal rings with three- and four-fold coordinated atoms. However, comparison with experimental evidence,[11] seems to indicate that this kind of structure is not present in real samples. In fact, this structure was obtained under the assumption that the dehydrogenation and lost of the head and tail groups of the BPs takes places simultaneously, which is probably a quite unrealistic situation. Therefore, we have considered the more realistic scenario of a BPT-SAM on Au(111) with the 2√3x√3 periodicity observed in STM experiments. We considered a monolayer of BPT molecules with a structure similar to the experimental one, but a slightly larger density, to allow for commensuration with a 2√3x2√3 supercell of Au(111) (each supercell contains four BPT molecules). Relaxing the structure after dehydrogenation of the BPTs, we observe that the molecules interact covalently to spontaneously form small "graphene-like" nanoflakes. Despite the increase of the Au-S distances the adsorption of these nanographenes on the Au(111) surface is still considerably stable. This fact, together with the limitations imposed by forcing commensurability in our periodic supercell calculations avoid the expected cross-link formation between these "graphene-like" nanoflakes and, thus, many carbon atoms in this optimized configuration remain still unsaturated. In any case, considering the limitations in our calculations, we believe that the graphene nanoflakes are a very plausible building block for the structure of the experimentally irradiated BPT-SAM adsorbed on Au(111) surface. Further effort is required to study the bonding between these graphene nanoflakes and the properties of the resulting extended structure.

**Acknowledgements**

We acknowledge support from Basque Departamento de Educación, UPV/EHU (Grant No. IT-366-07), the Spanish Ministerio de Educación y Ciencia (Grant No. FIS2007-66711-C02-00) and the ETORTEK research program funded by the Basque Departamento de Industria and the Diputación Foral de Guipúzcoa.